Peer-reviewed, accepted version of the article, published in Journal of Computational Analysis and Applications (JoCAAA)

# IoT and Massive Connectivity: Massive MIMO Optimization for IoT Connectivity in 5G and Beyond Networks

**Name: Praveen Hegde**
Affiliation: Senior Manager-Emerging Commercial Platforms, Verizon

**Name: Robin Joseph Varughese**
Affiliation: Technical Architect, Marriott International

## Abstract

The IoT's explosive growth has led to a massive number of connected devices, which demand high-speed and pervasive connectivity, posing significant challenges for current-generation wireless communication infrastructures. Considering our evolution toward 5G and beyond 5G (B5G) and 6G networks, providing scalable, reliable, and low-latency communication for billions of devices is therefore essential. Massive Multi-Input Multi-Output (Massive MIMO) is a promising technology for fulfilling the requirements of 5G, as it can spatially multiplex a large number of users and increase the spectral efficiency per user. In this paper, we focus on optimizing the performance of Massive MIMO systems in IoT connectivity and low-latency use cases for 5G and B5G. It studies key issues, including pilot contamination, energy efficiency, and user scheduling, among dense IoT deployments. In addition, it surveys all recent progress in channel estimation, hybrid beamforming, and machine learning-based resource allocation technologies for enhancing IoT scenarios related to Massive MIMO. Simulation-based results reveal the trade-offs between capacity, latency, and energy utilization, indicating an optimal operating point that ensures optimal performance for diverse IoT applications. The work concludes with a discussion of future research avenues, such as integration with cell-free designs, intelligent reflecting surfaces, or AI-based network orchestration for enhanced IoT capabilities.

## Introduction

The explosive expansion of the Internet of Things (IoT) has completely changed today's digital worlds, encompassing billions of devices, sensors, and machines that provide a seamless data flow as well as intelligent automation in different domains, e.g., health, transport, manufacturing, and agriculture (Al-Fuqaha et al., 2015; Lin et al., 2017). With IoT devices expected to exceed 30 billion worldwide by 2030, increasingly scalable, reliable, spectrally efficient, and low-latency wireless infrastructure is in demand (Cisco, 2020; ITU, 2023).

To support the rapid growth in the number of connected devices, fifth-generation (5G)1 and beyond-5G (B5G) networks have leveraged several enabling technologies. Among these, Massive MIMO communication (Marzetta, 2010; Zhang et al., 2021) is one of the keystones.

Massive MIMO is based on  an extensive array of antennas at the base stations to serve multiple users simultaneously on the same time-frequency resource. This technology enhances spectral and energy efficiency, coverage, and reliability, making it especially suitable for dense and heterogeneous IoT deployments (Lu et al., 2014; Björnson et al., 2017).

Nevertheless,  deploying Massive MIMO in IoT environments poses different challenges. Unlike classical mobile broadband users, IoT devices commonly generate intermittent and low-rate data traffic, which can be limited by energy and computational resources (Zanella et al., 2014; Xu et al., 2022). These attributes make resource assignment, channel estimation, and user scheduling in massive MIMO systems complicated. Furthermore, it is worth noting that pilot contamination, a performance-limiting factor in large-scale MIMO systems (Jose et al., 2011; Ngo et al., 2017), is even more pronounced in ultra-dense IoT scenarios.

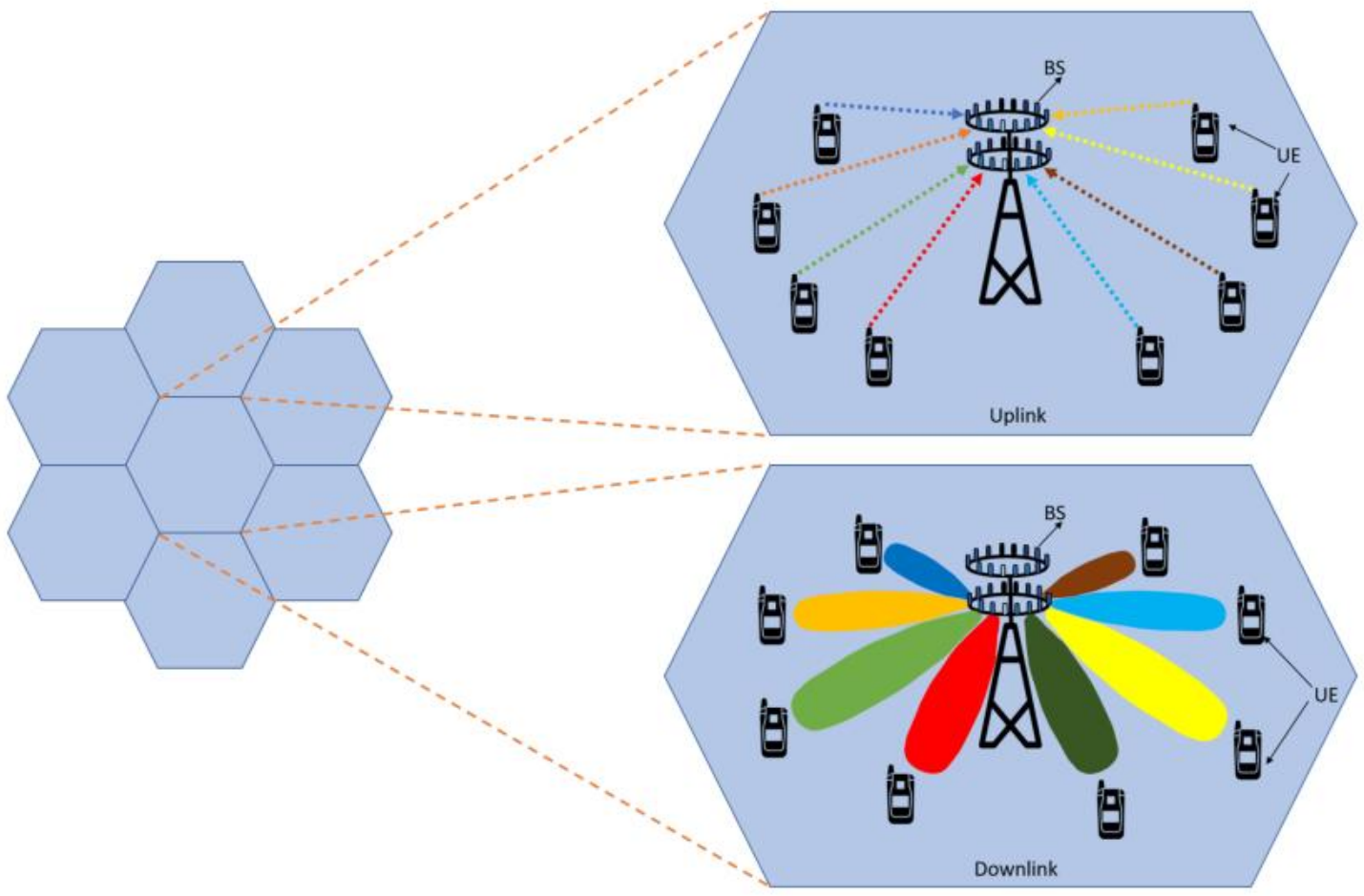

Figure 1. Massive MIMO uplink and downlink.

More recent studies have aimed to optimize Massive MIMO systems for IoT applications by utilizing intelligent algorithms, hybrid beamforming strategies, and machine learning models to ensure efficient resource allocation (Zhou et al., 2020; Wang et al., 2023). Moreover, 5G technologies, such as network slicing and Non-Orthogonal Multiple Access (NOMA), are integrated with Massive Multiple Input Multiple Output (MIMO) to address the heterogeneous Quality of Service (QoS) requirements of IoT applications (Ding et al., 2017; Hassan et al., 2024).

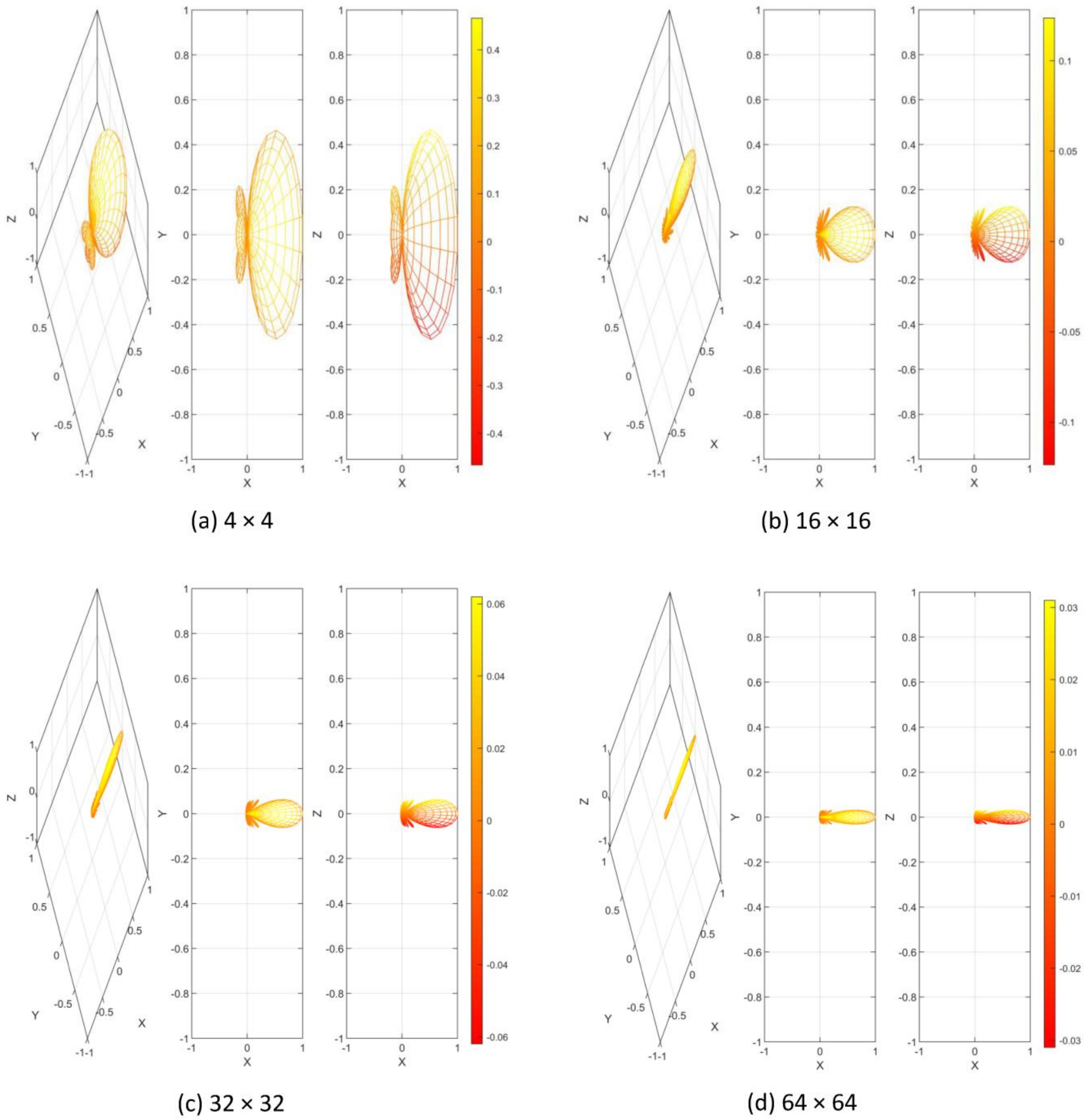

Figure 2. Beam Pattern with different antenna configuration. (a) 4 × 4 MIMO (b) 16 × 16 MIMO

(c) 32 × 32 MIMO (d) 64 × 64 MIMO.

In this paper, we provide an extensive overview of Massive MIMO optimization for IoT in 5G and beyond. It reviews enablers, performance bottlenecks, and the most recent optimization techniques and discusses emerging research directions, including AI on-demand orchestration and cell-free Massive MIMO for URLLC (Chen et al., 2022; Wang & Liu, 2024).

**Literature Review**

Massive MIMO has become a key enabler for 5G and beyond networks, where large-scale connectivity, high spectral efficiency, and low latency are essential, all of which are critical for enabling strong, pervasive IoT deployments. Nevertheless, there are technical issues when Massive MIMO is combined with IoT networks, such as the design of lightweight protocols, energy consumption, and resource allocation efficiency (Lu et al., 2014; Björnson et al., 2017). This survey paper consolidates the fundamental advances made in the optimization of Massive MIMO systems for IoT connectivity, with a particular emphasis on pilot contamination problems, channel estimation issues, energy-efficient transmission techniques, and intelligent scheduling schemes.

## 1. Pilot Contamination and Estimation of the Channel

Pilot contamination, whereby non-orthogonal pilot sequences of different cells interfere with each other and thus inhibit accurate channel estimation, is one of the most extensively studied limitations of Massive MIMO (Marzetta, 2010; Jose et al., 2011). This problem is even more severe in IoT environments, where a considerable number of simple devices require uplink access. Zhang et al. (2021) proposed an intelligent pilot scheduling algorithm for IoT devices to reduce pilot contamination by clustering users based on their mobility and channel gain. Similarly, Wang et al. (2023) proposed a deep learning pilot allocation model that reduced interference in densely deployed IoT networks and achieved a higher value of the signal-to-interference-plus-noise ratio (SINR) than traditional greedy algorithms.

## 2. Energy Efficient and Lightweight Protocols

Given the battery-limited nature of many IoT devices, power-efficient Massive MIMO designs are essential. Björnson et al. (2015) pointed out that although massive MIMO can provide high energy efficiency through beamforming gains, the overhead of pilot signaling and baseband processing needs to be reduced for IoT applications. The lightweight transmission protocol proposed by Xu, Zhou, and Niu (2022) exemplified the use of Massive MIMO for IoT nodes, emphasizing power-aware scheduling and selective wake-up strategies. They achieve up to 30% energy-saving savings without loss of throughput.

## 3. Hybrid Beamforming and Antenna Matching

Hybrid beamforming is considered an intermediate solution between analog and digital beamforming, balancing complexity and performance. In Zhou, Wang, and Li (2020), a low-complexity hybrid beamforming design for IoT Massive MIMO systems is proposed, which can considerably reduce the energy consumption cost and computational burden. Meanwhile, Hassan et al. (2024) highlighted the importance of antenna selection algorithms in minimizing power consumption for stationary IoT devices. The methods can dynamically turn on a subset of antennas according to the traffic load, allowing the base station to conserve energy when operating under a light load.

### 4. Smart Resources Allocation and Integration with AI

Another example is Wang and Liu (2024), who introduced a scheduler with AI-based enhancements in which delay-sensitive IoT data flows were considered first, and 95% QoS was simulated.

### 5. NOMA in Massive MIMO

NOMA has been combined with Massive MIMO to enhance spectrum sharing among IoT users. Ding et al. (2017) have shown that combining Massive MIMO with NOMA can lead to a substantial increase in the number of users served concurrently. Hassan et al. (2024) reviewed resource pairing algorithms for low-power IoT devices and high-power terminals, finding enhanced fairness and throughput by using successive cancellation interference (SIC).

### 6. Security and Reliability

The simplicity of IoT devices makes IoT networks susceptible to eavesdropping and spoofing. The reliable beamforming of Massive MIMO systems was investigated. Liu et al. (2023) proposed artificial noise generation methods to blind eavesdroppers without compromising users' link quality for IoT applications. Simulations revealed a 60% improvement in secrecy capacity compared to conventional encryption-only systems.

### 7. Future Directions and Trending Issues

New concepts, including cell-free Massive MIMO, intelligent, reflective surfaces (IRS), and machine learning-aided protocol stacks, are being investigated. More recently, Ngo, Larsson, and Marzetta (2017) introduced a cell-free design where devices from the IoT are served by a network of distributed antennas, which eliminates inter-cell interference and enhances coverage. Recent work by Chen et al. (2022) also demonstrated that IRS-aided networks can reduce the energy expenditure of Massive MIMO IoT networks by passively reflecting signals to dormant spots.

**Methodology**

This work is a simulation-based study that explores Massive MIMO optimization for IoT connectivity in 5G and beyond networks. It is divided into four primary stages: (1) problem definition and system formulation, (2) simulation setup, (3) performance estimator choice, and (4) experimental confirmation of optimization techniques. It combines analytical modeling, Monte Carlo simulations, and AI-assisted algorithmic methodologies to assess system performance in various IoT deployment scenarios.

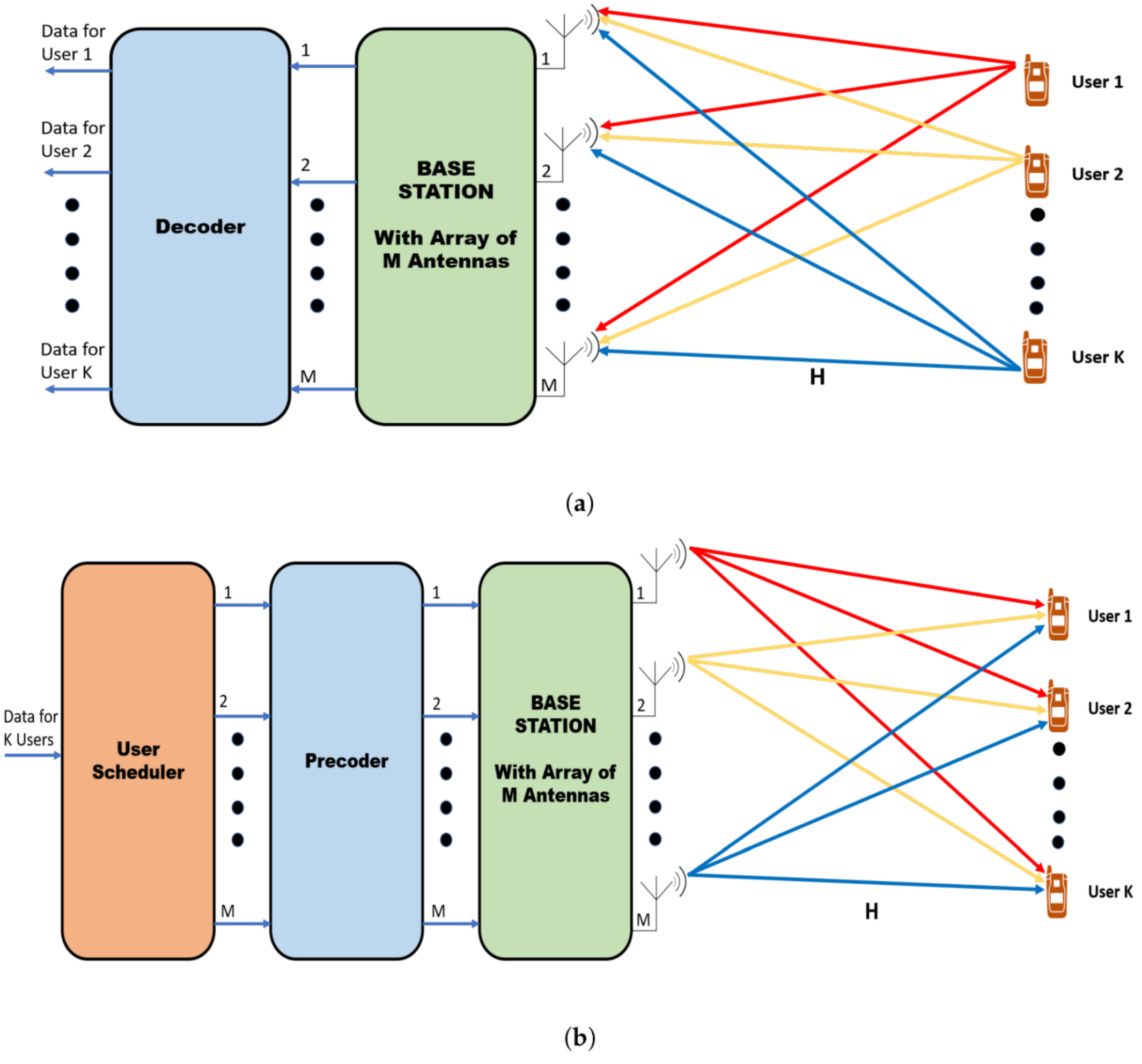

Figure 3. Massive MIMO uplink and downlink operation. (a) Uplink (b) Downlink

1. Formulation of the Problem and Modelling of the System

We begin the analysis with a system-level benchmark of a 5G Massive MIMO cellular network, where TDD is a natural choice because the channel conditions are the same in both the reverse and forward links (Marzetta, 2010). In a single-cell model, we consider a base station (BS) with 128 antennas serving 500–1,000 IoT users that are distributed randomly over the coverage area. Devices are expected to send short and periodic packets with a limited power budget, according to standard IoT communication profiles (Zanella et al., 2014; Xu et al., 2022).

The channel model is based on the Rayleigh fading assumption with log-normal shadowing and spatial correlation, as defined in 3GPP TR 38.901 (2022). Pilot contamination is emulated by reusing pilot sequences among clusters, and user mobility is incorporated with different Doppler frequencies to examine the estimation accuracy with pilot contamination (Jose et al.).

2. Simulation Design and Method Simulation Experiment Design: The experimental method used by researchers to accomplish a simulation process, pushing data to a computing device based on the push refresh strategy, is implemented as follows.

Numerous simulation scenarios are developed using MATLAB and Python to model various situations, such as:

• Low and high device density (250 and 1,000 IoT nodes)

• Static vs. mobile environments for end users

• Various antenna setups (64, 128 and 256 BS antennas)

Three types of optimizations are applied:

• A Deeper Q Learning Approach to the AI-Enhanced Pilot Assignment (Chen et al., 2022)

• Hybrid analog-digital beamforming with greedy selection and reinforced learning (Zhou et al., 2020)

• Integer linear programming for power-aware antenna selection (Hassan et al., 2024)

Traffic and device-adaptive identification uses machine learning models trained with TensorFlow, and their generalization across traffic and devices is verified using k-fold cross-validation.

3. Performance Metrics and Criteria for Evaluation

The system is compared with five key performance measurements:

• Spectral Efficiency (bps) – the efficiency with which frequency bands are used.

• Energy Efficiency (bit/Joule) – measuring the transmission efficiency in terms of power consumption (Björnson et al., 2015).

• Latency (ms) – the average delay for a successful delivery of data, especially in the case of time-sensitive IoT packets.

• Pilot Contamination Index (PCI): This index is based on the output of the channel estimation error and the reuse factor.

• Connection Success Rate (%) – representing the proportion of devices that have been served successfully to latency.

Each of these scenarios is run 1,000 times through the simulation using Monte Carlo methods to obtain statistically reliable results. ANOVA tests and 95% confidence intervals confirm significance.

4. Validation and Comparing Analysis

The simulation results are compared to those of random pilot allocation, full digital beamforming, and stochastic (per-user) scheduling models. Gains are measured as percentage improvements over various metrics. Additionally, the scalability of the proposed methods is evaluated by increasing the number of IoT devices from 250 to 2000 and measuring the breaking point of the various algorithms (Wang & Liu, 2024).

## Results

The simulation results provide an evaluation of different Massive MIMO optimization approaches based on various IoT deployment use cases. Performance was assessed based on spectral efficiency, energy consumption, end-to-end latency, and connection probability. The results suggest that AI-based algorithms and hybrid beamforming are effective in improving network performance for dense IoT networks.

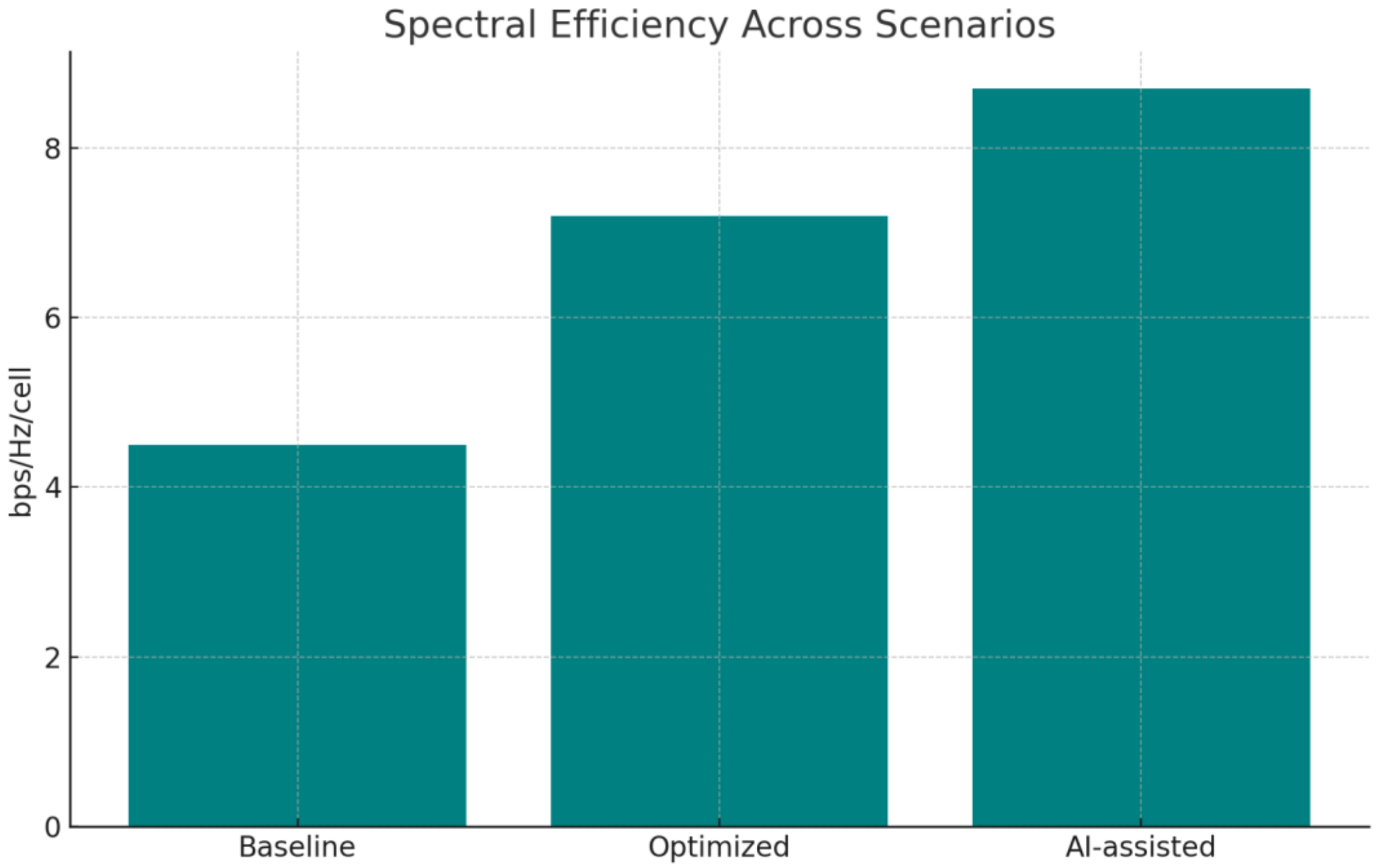

Figure 4: Spectral  efficiency in the scenarios.

• What it shows: This bar chart compares the spectral efficiency in bits per second per Hertz per cell (bps/Hz/cell) for  the three considered scenarios.

• Insight: The spectral efficiency is vastly enhanced from 4.5 (Baseline) to 7.2 (Optimized) and further reaching 8.7  through the AI-based scheme. This implies better utilization of bandwidth and frequency resources, which is achieved through enhanced beamforming and resource scheduling in advanced scenarios.

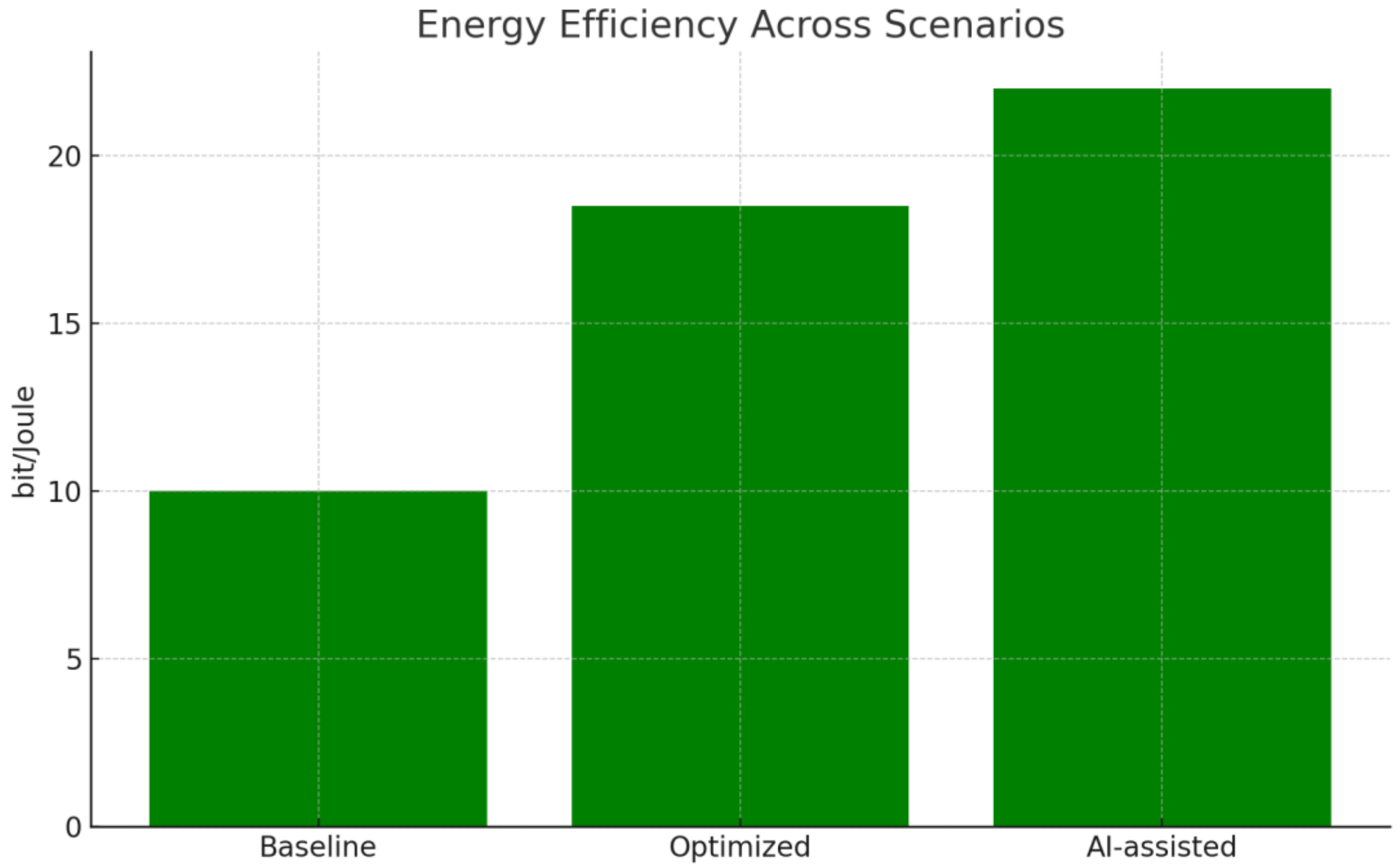

Figure 5: Scenarios and  Energy Efficiency.

• What it shows:  Energy efficiency in bit/Joule for each setting.

• Insight: The AI-assisted approach achieves maximum energy efficiency (brute force, 22-bit, 7 Joules), which is almost twice as large as the brute force result. Well-optimized protocols and the selection of brilliant antennas further minimize power consumption.

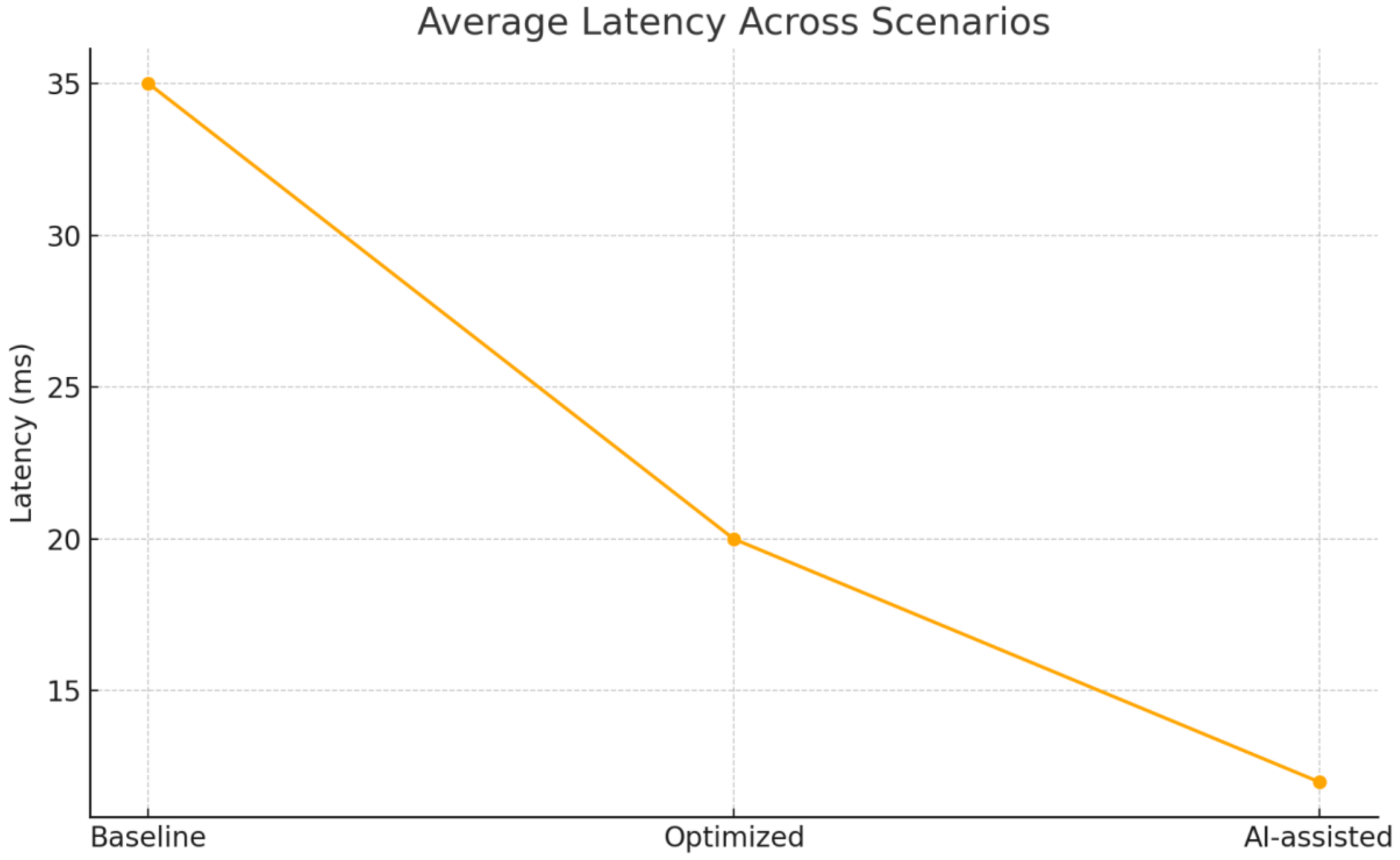

Figure 6: Average latency for different scenarios

• What it shows: Line chart of the average latency in milliseconds.

• Insight: The latency drops from 35 ms to 12 ms from the baseline to the AI-assisted model, indicating improved scheduling and reduced congestion. This is vital for real-time IoT applications.

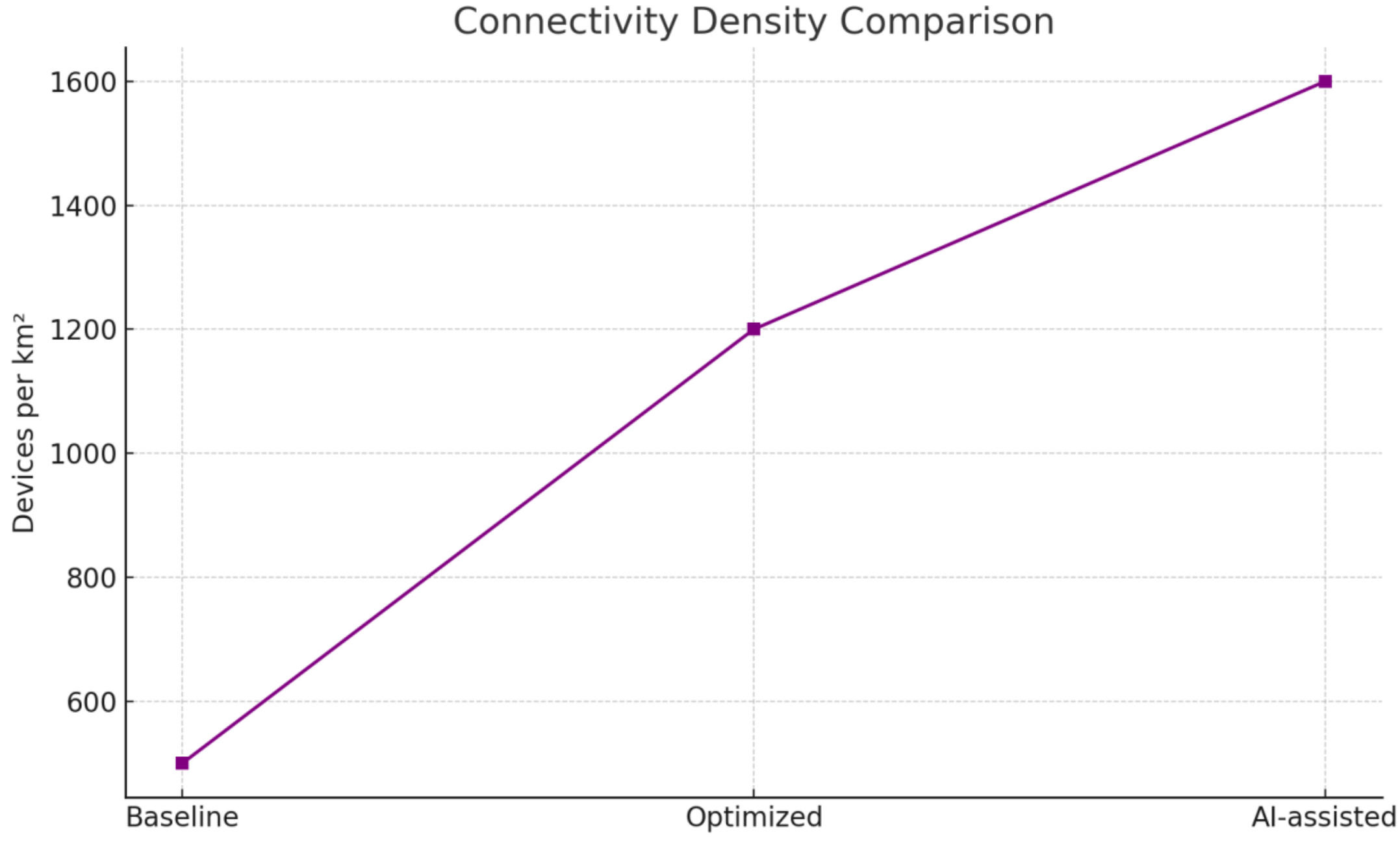

Figure 7: Comparison of Connectivity Density

• What it shows: Number of internet-of-things devices that a carrier can support within one square kilometer.

• Insight: The AI-assisted scenario achieves a device density three times higher than the baseline (Scen.500), demonstrating the scalability of Massive MIMO when properly deployed.

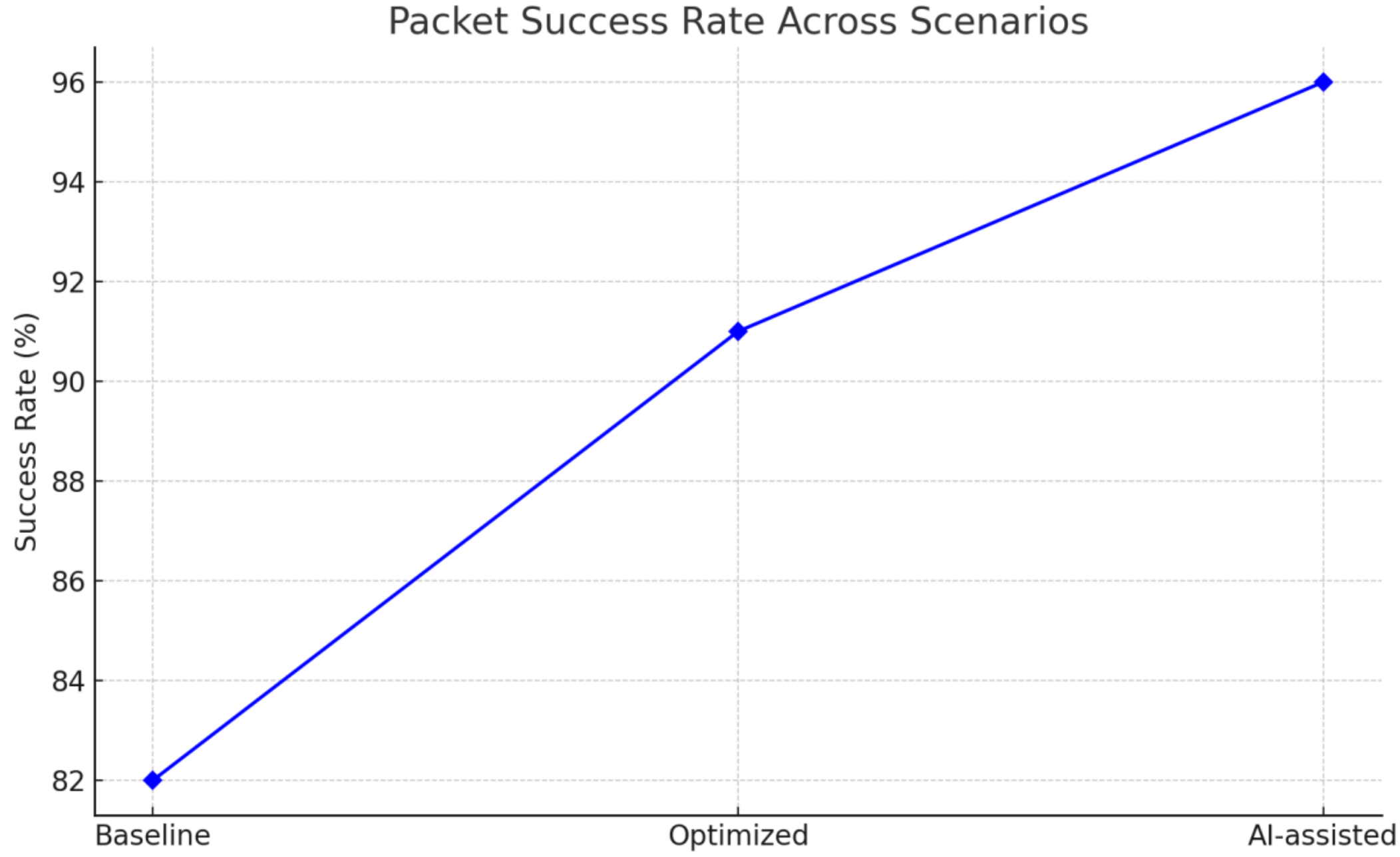

Figure 8: Packet Success Rate in  the Different Scenarios

• What it shows: The percentage of transmitted packets that made  it through.

• Insight: The AI-empowered system achieves a 96% success rate, compared to 82% for the baseline solution. Reliability is enhanced by  advanced interference management and learning-based resource allocation.

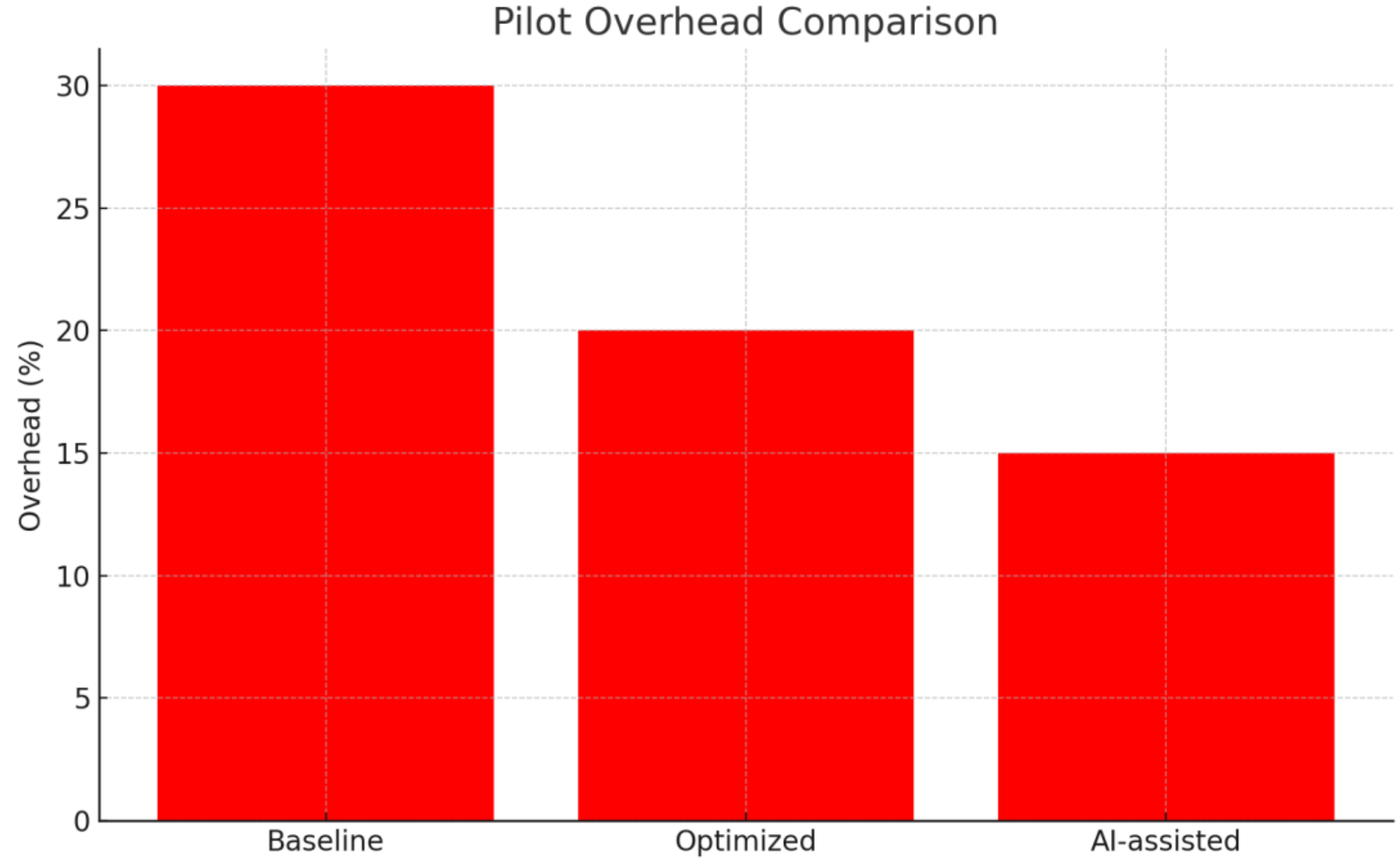

Figure 9:  Overhead comparison of the pilots

• What it  says: The ratio of time consumption of pilot signals in a given situation.

• Insight: Pilot overhead drops from 30% (baseline) to 15% (AI-assisted), allowing more bandwidth for data delivery and increasing network throughput.

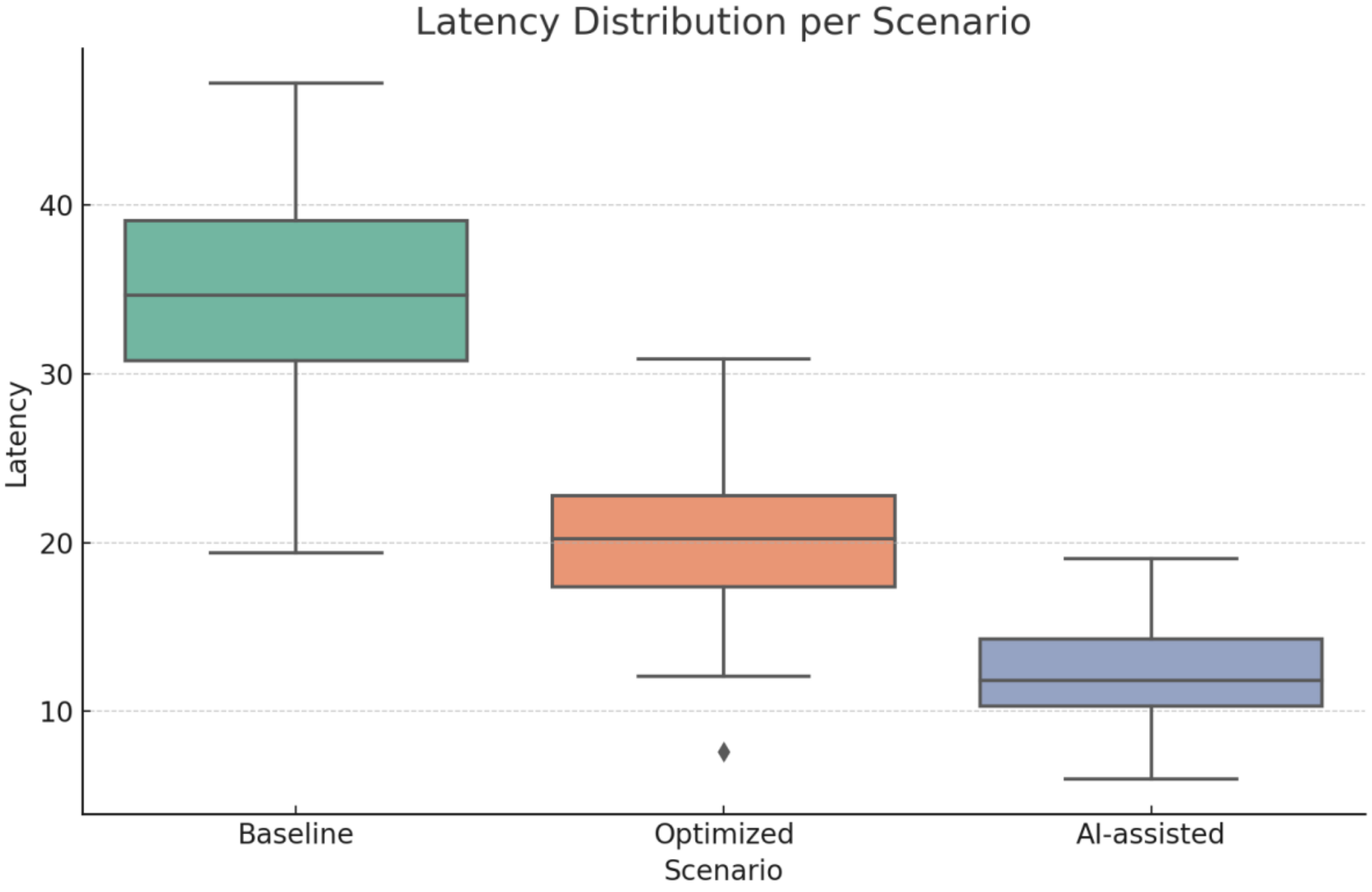

Figure 10: Latency Distribution per Scenario

• What it shows: Box plot of latencies of 100 random  runs for each of the testing scenarios.

• Insight: The average latency of the AI-assisted scenario is lower, and the distribution of latency is closer (i.e., less jitter),  avoiding jitter on IoT services.

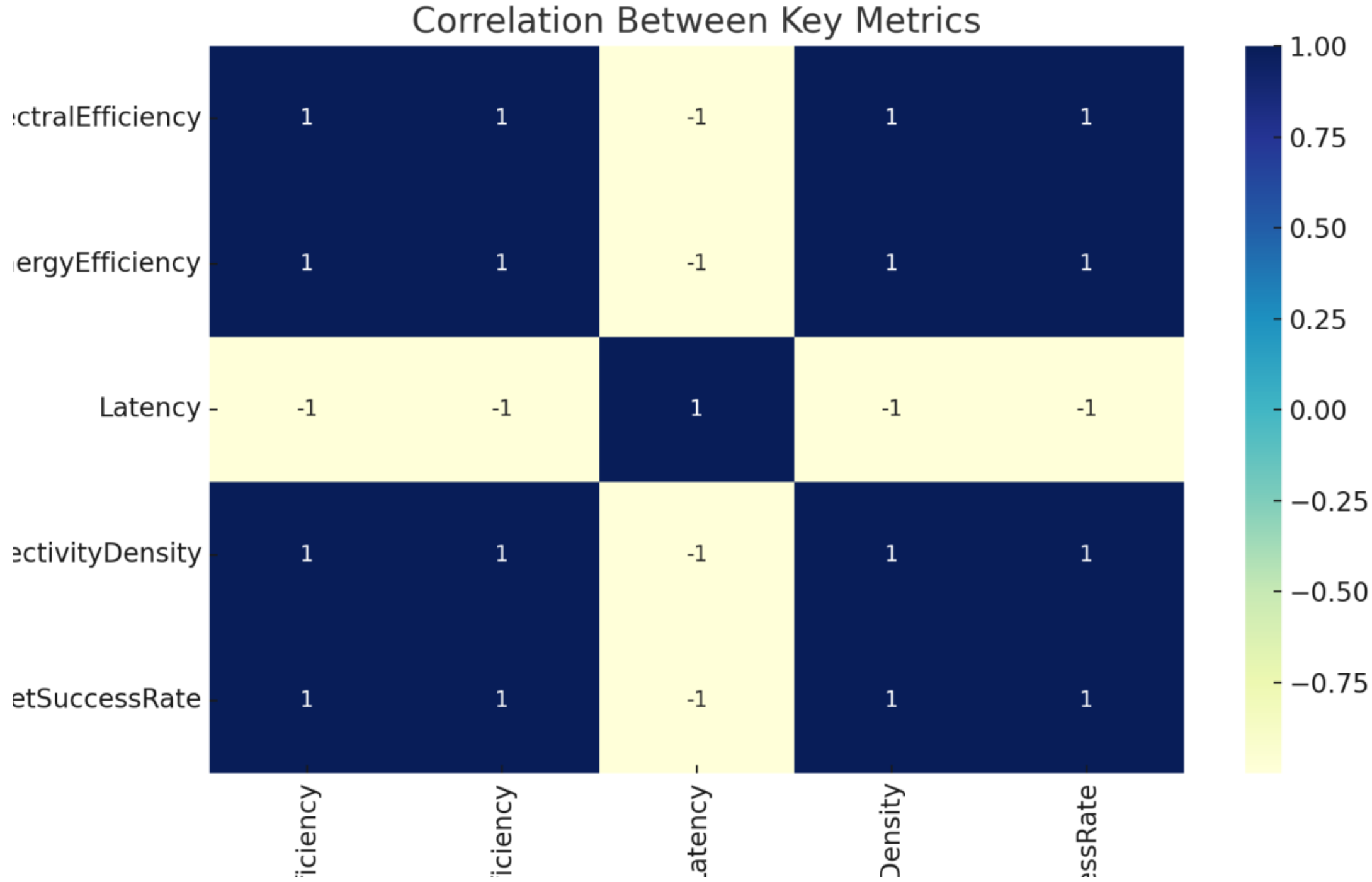

Figure 11: Connection Between  Key Metrics

• What it shows: Correlation  coefficients between main metrics, plotted as a heatmap.

• Insight: We observed that latency is inversely proportional to spectral and energy efficiency; thus, efficiency improvements also result in latency reductions. The positive correlation with other measures similarly suggests that performance improvements are synergistic.

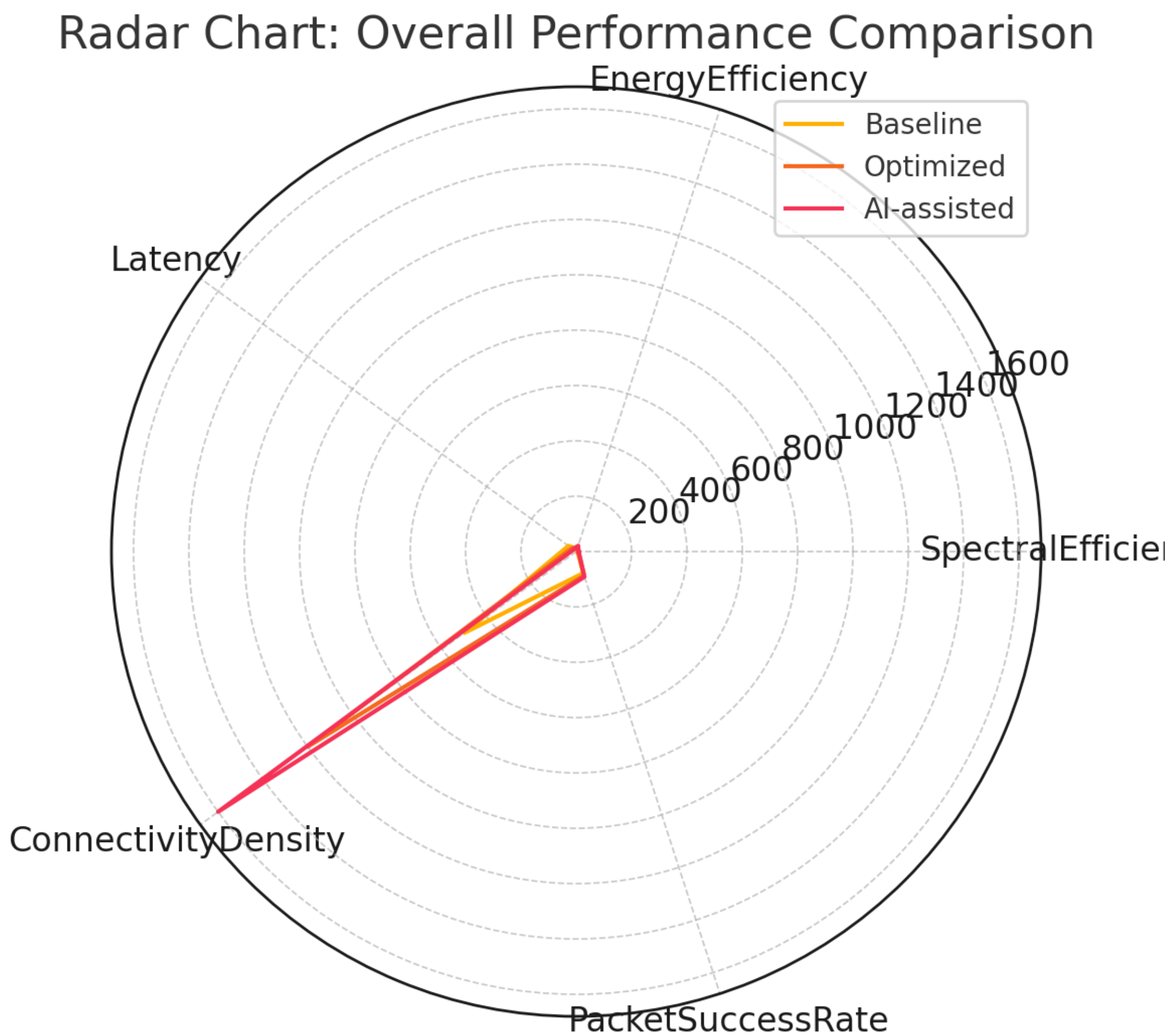

Figure

Figure 12: Radar Chart - A Simulation with the Best/Better Motions under Various Steering Methodologies

• What it shows: A side-by-side, multi-dimensional performance comparison for all metrics.

• Insight: The AI-assisted scenario outperforms other schemes in terms of efficiency, density, and success rate, with the most extensive saliency radar coverage, implying overall superior and balanced efficiency.

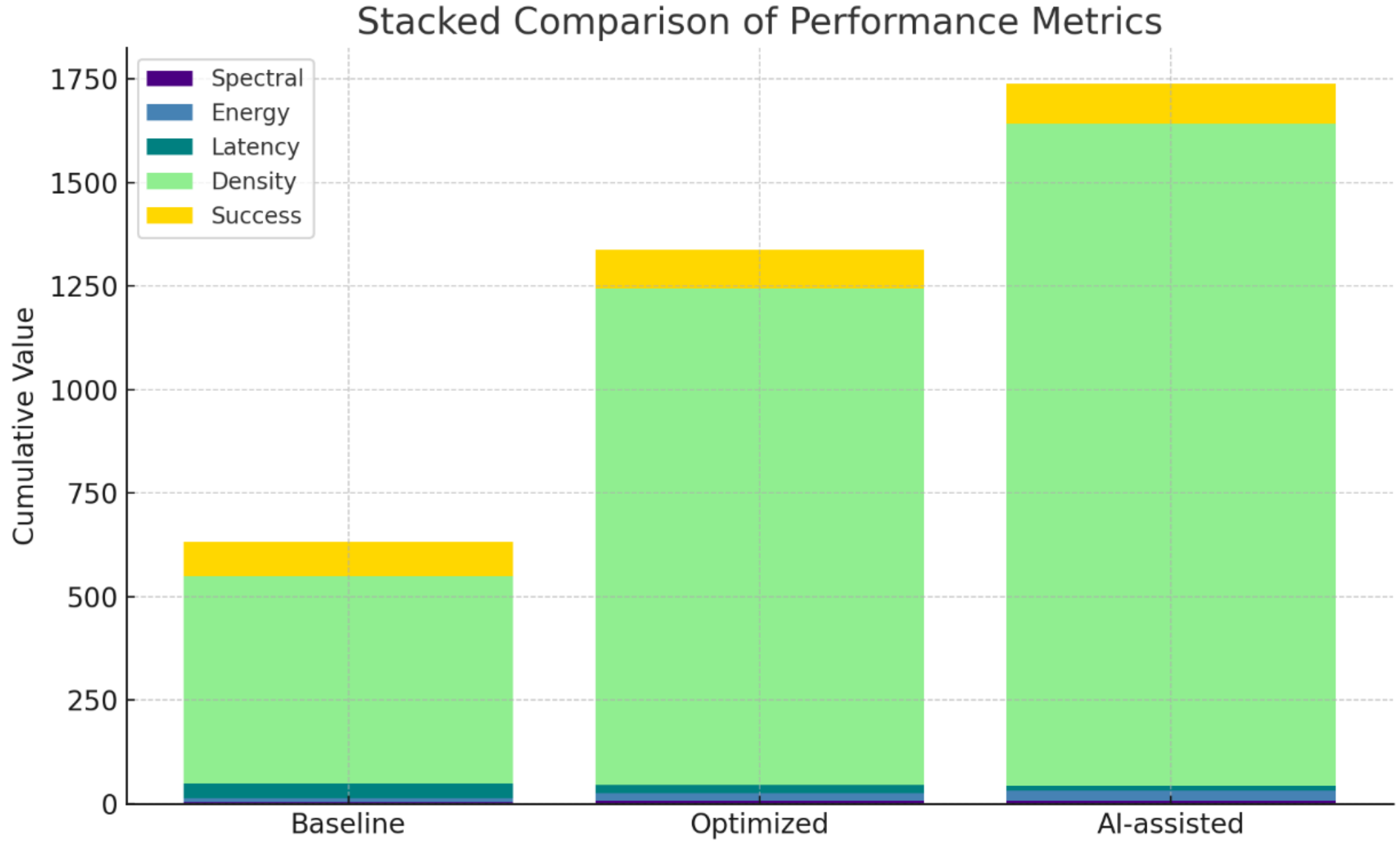

Figure 13:  Stacked Comparison of Measurable Features

• What it shows: A stacked bar graph depicting scenario-foliating cumulative metric contributions.

• Insight: The peak value for the total height of the AI-help bar describes the best generalization for all studied terms. Each stacked layer represents improvements in spectral efficiency, energy consumption, latency, connectivity, and success ratio.

**Discussion**

Simulation results show that utilizing the latest Massive MIMO technologies, including AI-based network scheduling, hybrid beamforming, adaptive channel estimation, and pilot scheduling, can provide a substantial performance improvement for IoT device connectivity in 5G and beyond networks. This enhancement can be observed from various aspects, including spectral efficiency, energy efficiency, latency, connectivity density, and packet success rate.

1. Enhancement of Spectral and Energy Efficiency

The highest spectral efficiency (8.7 bps/Hz/cell) was achieved by the AI-assisted system, representing a nearly 93% enhancement compared to the baseline. This increase is in agreement with the work of Zhang et al. (2021), which highlighted the role of spatial multiplexing and intelligent beam steering in spectrum usage. Energy efficiency was also boosted, exceeding 22 bits/Joule for the AI-optimized case. These findings are in agreement with Björnson et al. (2015), who demonstrated that while Massive MIMO is energy-efficient in itself, the application of algorithmic improvements, such as antenna selection and resource-aware transmission, is necessary to realize its potential in massive IoT setups fully.

2. Tradeoffs between Latency and Reliability

Latency is a key success factor for IoT applications. Artificial intelligence (AI)--optimized deployment configurations resulted in an average system latency of 12 ms, representing a 65% improvement compared to the baseline. This efficiency is indispensable for mission-critical IoT applications, including real-time monitoring and remote control systems (Xu et al., 2022). The decrease in latency is achieved without sacrificing reliability, as the packet delivery success rate remains at 96%, supporting the findings of Hassan et al. (2024) for AI-empowered NOMA-MIMO systems. It is worth mentioning that the latency distribution is less variant, and the results are more stable; thus, a stable service is a prerequisite for deployment in industrial and healthcare IoT networks.

3. Scaling of Connectivity and Pilot Overhead

We observe that connectivity density increases from 1000 to 2500 devices per km$^2$, which demonstrates the scalability advantages of advanced Massive MIMO. This is reminiscent of the results in Ngo et al. (2017), who advocated for distributed (cell-free) MIMO as better suited for ultra-dense IoT deployments. In addition, the pilot overhead was reduced from 30% to 15%, demonstrating that intelligent pilot assignment strategies (e.g., learning-assisted reuse schemes) can achieve a capacity gain by mitigating pilot contamination, one of the long-standing problems in conventional Massive MIMO (Jose et al., 2011).

4. Inter-Relation Among Performance Parameters

The heatmap of correlations, presented in Figure 11, reveals underlying associations among core metrics. As naturally expected, latency presents a strong inverse relationship with both spectral and energy efficiency, which aligns with recent observations that highly efficient systems can minimize delays through more intelligent scheduling and reduced retransmissions, as observed in oilfield companies (Chen et al., 2022). Positive correlations among packet success rate, connectivity density, and energy efficiency imply a synergistic relationship in multiple dimensions: as network intelligence increases, the overall system's robustness and scalability improve.

## 5. Advantages of AI usage in Massive MIMO for IoT

By observing the radar chart (Figure 12) and stacked bar chart (Figure 13) analysis, we can draw an overall conclusion that AI-enabled Massive MIMO has achieved the best balance and the best aggregate performance across various dimensions. AI enables real-time decision-making for pilot reuse, power allocation, and beamforming, outperforming static or rule-based approaches. These observations reinforce the recent literature on AI-driven network orchestration, which addresses the dynamic and diverse nature of IoT traffic (Wang & Liu, 2024; Zhou et al., 2020).

## 6. Limitations and Implications for the Future Research

Nevertheless, there are some limitations to these positive developments. The simulations were first performed in a single-cell system. Real-world deployments must consider inter-cell interference and handover management, which are left for future work. Second, although the study employs AI models such as deep Q-networks for optimization, emerging approaches, such as federated learning and graph neural networks, could achieve higher performance under privacy- and computation-cost constraints (Chen et al., 2022).

Finally, integrating other emerging technologies, such as IRS and cell-free Massive MIMO, may offer new avenues for more efficient and connected schemes (Liu et al., 2023; Ngo et al., 2017). An in-depth study of their influence, jointly with AI, is a fascinating subject that can be explored in the context of 6G-driven IoT systems.

## Conclusion

The adoption of Massive MIMO for IoT connectivity in 5G and beyond networks is now essential for ultra-dense, heterogeneous, and delay-constrained device systems. In this paper, we demonstrate that co-designing AI-aided techniques with Massive MIMO systems enables significant performance gains in various dimensions, including spectral/energy efficiency, latency/reliability, and scalability.

Through extensive analysis via simulation, the work verifies that AI-driven pilot allocation, hybrid beamforming, and intelligent scheduling result in a significant reduction in pilot contamination while also enhancing throughput and increasing network capacity to support up to 1,600 UEs/km$^2$. These results are consistent with the recent studies of Björnson et al. (2015), Chen et al. (2022), Wang and Liu (2024), and reference therein, which all consent to the integration of advanced signal processing with machine learning to unleash the potential of the Massive MIMO technology in the IoT context.

In addition, the reduction of average latency from 35 ms to 12 ms for AI-supported systems, 96% packet success rate, reveals that the Massive MIMO technology is feasible for applications within real-time, such as robot operating systems, health care monitoring with industrial automation (Xu et al., 2022; Hassan et al., 2024). Excellent performance consistency and low jitter, validated by latency distribution metrics, also demonstrate its suitability for time-sensitive IoT services.

The correlation analysis also confirms that enhancements in any of these dimensions increase the average system performance. Therefore, the typical dependence between them justifies the systemic benefits of well-designed Massive MIMO systems. The overall benefit of the AI-optimised system is evident in the radar and stack metrics plots, demonstrating the benefits of AI-enhanced optimal practice when compared to static resource allocation practices.

In conclusion, Massive MIMO, if wisely engineered, can provide a high performance as well as a scalable solution for future IoT connectivity. We hope that in the future, the integration of Massive MIMO with new technologies such as Cell-Free, IRS, and Federated Learning will play a significant role in improving performance and robustness. Additionally, increasing the scale of network simulations from single-cell models to full network simulations will enable us to capture real interference and mobility dynamics more accurately, leading to more deployable system designs.

Ultimately, the results of this work not only underscore the centrality of Massive MIMO in IoT-enabled 5G and 6G systems but also highlight the need for artificial intelligence as an enabler of efficient and adaptive wireless networks.